\documentclass[twocolumn,conference]{IEEEtran}
\usepackage[T1]{fontenc}
\usepackage[latin9]{inputenc}
\usepackage{units}
\usepackage{amsmath}
\usepackage{amsthm}
\usepackage{amssymb}
\usepackage[unicode=true,
 bookmarks=true,bookmarksnumbered=true,bookmarksopen=true,bookmarksopenlevel=1,
 breaklinks=false,pdfborder={0 0 0},pdfborderstyle={},backref=false,colorlinks=false]
 {hyperref}
\hypersetup{pdftitle={Your Title},
 pdfauthor={Your Name},
 pdfpagelayout=OneColumn, pdfnewwindow=true, pdfstartview=XYZ, plainpages=false}

\makeatletter
\theoremstyle{plain}
\newtheorem{thm}{\protect\theoremname}
\theoremstyle{definition}
\newtheorem{defn}[thm]{\protect\definitionname}
\theoremstyle{plain}
\newtheorem{prop}[thm]{\protect\propositionname}
\theoremstyle{definition}
\newtheorem{example}[thm]{\protect\examplename}
\theoremstyle{plain}
\newtheorem{conjecture}[thm]{\protect\conjecturename}
\theoremstyle{plain}
\newtheorem{lem}[thm]{\protect\lemmaname}
\theoremstyle{plain}
\newtheorem{cor}[thm]{\protect\corollaryname}

\usepackage[caption=false,font=footnotesize]{subfig}

\usepackage{epsfig}

\usepackage{pgf}
\usepackage{tikz}\usepackage{mathrsfs}
\usetikzlibrary{arrows}

\makeatother

\providecommand{\conjecturename}{Conjecture}
\providecommand{\corollaryname}{Corollary}
\providecommand{\definitionname}{Definition}
\providecommand{\examplename}{Example}
\providecommand{\lemmaname}{Lemma}
\providecommand{\propositionname}{Proposition}
\providecommand{\theoremname}{Theorem}

\begin{document}

\title{Quantum Information on Spectral sets}

\author{\IEEEauthorblockN{Peter~Harremo{\"e}s}\IEEEauthorblockA{Niels Brock\\
Copenhagen Business College\\
Copenhagen\\
E-mail: harremoes@ieee.org}}
\maketitle
\begin{abstract}
For convex optimization problems Bregman divergences appear as regret
functions. Such regret functions can be defined on any convex set
but if a sufficiency condition is added the regret function must be
proportional to information divergence and the convex set must be
spectral. Spectral set are sets where different orthogonal decompositions
of a state into pure states have unique mixing coefficients. Only
on such spectral sets it is possible to define well behaved information
theoretic quantities like entropy and divergence. It is only possible
to perform measurements in a reversible way if the state space is
spectral. The most important spectral sets can be represented as positive
elements of Jordan algebras with trace 1. This means that Jordan algebras
provide a natural framework for studying quantum information. We compare
information theory on Hilbert spaces with information theory in more
general Jordan algebras, and conclude that much of the formalism is
unchanged but also identify some important differences.
\end{abstract}

\section{Introduction}

Although quantum physics has existed for more than 100 years there
are still many simple fundamental questions that remain unanswered.
One of the main questions is why complex Hilbert spaces are used.
Although the complex numbers are extremely useful for doing computations
we do not have direct observations of these numbers. Observables are
represented by selfadjoint operators and such operators can be multiplied
but the multiplication leads to operators that are not selfadjoint
and cannot be observed. Therefore Pascual Jordan introduced what is
now known as Jordan algebras with a different product than the usual
matrix product. Still, his product is not well physically motivated.
In this paper we give some general definitions of some information
theoretic quantities and demonstrate that they are only well behaved
on convex sets that can be represented on Jordan algebras. Therefore
the Jordan algebras appear to be the correct formalism for doing quantum
information. From a computer science point of view the quantum algorithms
should be run in Jordan algebras, but for building a quantum computer
we need a physical implementation of the algorithm. Appearently the
physical world prefer Jordan algebras associated with complex Hilbert
spaces so a general algorithm on a Jordan algebra should be implemented
on complex Jordan algebras. This is possible in most cases but may
increase the number of gate operations by a factor.

\section{Structure of the state space}

Our knowledge about a system will be represented by a state space.
I many cases the state space is given by a set of probability distributions
on the sample space. In such cases the state space is a simplex, but
it is well-known that the state space is not a simplex in quantum
physics. In order to cover applications in physics we need a the more
general notion of a state space as defined in \cite{Holevo1982}.

\subsection{The state space and the positive cone}

Before we do any measurement we prepare our system. Let $\mathcal{P}$
denote the set of preparations. Let $p_{0}$ and $p_{1}$ denote two
preparations. For $t\in\left[0,1\right]$ we define $\left(1-t\right)\cdot p_{0}+t\cdot p_{1}$
as the preparation obtained by preparing $p_{0}$ with probability
$1-t$ and $p_{1}$ with probability $t$. A measurement $m$ is defined
as an affine mapping of the set of preparations into a set of probability
measures on some measurable space. Let $\mathcal{M}$ denote a set
of feasible measurements. The state space $\mathcal{S}$ is defined
as the set of preparations modulo measurements. Thus, if $p_{1}$
and $p_{2}$ are preparations then they represent the same state if
$m\left(p_{1}\right)=m\left(p_{2}\right)$ for any $m\in\mathcal{M}.$

In statistics the state space equals the set of preparations and has
the shape of a simplex. The symmetry group of a simplex is simply
the group of permutations of the extreme points. In quantum physics
the state space has the shape of the density matrices on a complex
Hilbert space and the state space has a lot of symmetries that a finite
simplex does not have. For simplicity we will assume that the state
space is a finite dimensional convex compact space.

From now on we consider the situation where our knowledge about a
system is given by an element in a convex set. These elements are
called \emph{states} and convex combinations are formed by probabilistic
mixing. States that cannot be distinguished by any measurement are
considered as being the same state. The extreme points in the convex
set are called \emph{pure states} and all other states are called
\emph{mixed states}. See \cite{Holevo1982} for details about this
definition of a state space. 

If $\mathcal{S}$ is a state space it is sometimes convenient to consider
the positive cone generated by $\mathcal{S}$ . The positive cone
consist of elements of the form $\lambda\cdot\sigma$ where $\lambda\geq0$
and $\sigma\in\mathcal{S}$. Elements of a positive cone can multiplied
by positive constants via $\lambda\cdot\left(\mu\cdot\sigma\right)=\left(\lambda\cdot\mu\right)\cdot\sigma$
and can be added as follows. 
\[
\lambda\cdot\rho+\mu\cdot\sigma=\left(\lambda+\mu\right)\cdot\left(\frac{\lambda}{\lambda+\mu}\rho+\frac{\mu}{\lambda+\mu}\sigma\right).
\]
  The convex set and the positive cone can be embedded in a real vector
space by taking the affine hull of the cone and use the apex of the
cone as origin of the vector space.

\subsection{Measurements on the state space}

Let $m$ denote a measurement on the state space $\mathcal{S}$ with
values in the set $A.$ Then $m\left(\sigma\right)$ is a probability
distribution on $A.$ Let $B\subseteq A.$ Then $m\left(\sigma\right)\left(B\right)$
is the probability of observing a result in $B$ when the state is
$\sigma$ and the measurement $m$ is performed. Then $m\left(\sigma\right)\left(B\right)\in\left[0,1\right]$
and $\sigma\to m\left(\sigma\right)\left(B\right)$is an affine mapping.
\begin{defn}
Let $\mathcal{S}$ denote a convex set. A \emph{test} is an affine
map from $\mathcal{S}$ to $\left[0,1\right]$ . 

The tests are building block for all measurements accoring to the
following proposition.
\end{defn}
\begin{prop}
Let $m$ denote a measurement on the convex set $\mathcal{S}$ with
values in the finite set $A.$ Then there exists tests $\phi_{b}:\mathcal{S}\to\left[0,1\right]$
such that for any $B\subseteq A$ we have
\begin{equation}
m\left(\sigma\right)\left(B\right)=\sum_{b\in B}\phi_{b}\left(\sigma\right).\label{eq:decomposition}
\end{equation}
\end{prop}
If the set $A$ is not finite we may have to replace the sum by an
integral. The trace of a positive element is defined by $\mathrm{tr}\left(\lambda\cdot\sigma\right)=\lambda$
when $\sigma\in\mathcal{S}$ so that states are positive elements
with trace equal to 1. We note that the trace restricted to the states
defines a test. Any tests can be identified with a positive functional
on the positive cone that is dominated by the trace. Note that if
a measurement $m$ is given by (\ref{eq:decomposition}) then the
tests satisfies
\[
\sum_{b\in A}\phi_{b}=\mathrm{tr}\,.
\]

\subsection{Improved Caratheodory theorem}

Let $x$ be an element in the positive cone such that 
\begin{equation}
x=\sum_{i=1}^{n}\lambda_{i}\cdot\sigma_{i}.\label{eq:SpecDecomp}
\end{equation}
where $\sigma_{i}$ are pure states. If $\lambda_{1}\geq\lambda_{2}\geq\dots\geq\lambda_{n}$
then the vector $\lambda_{1}^{n}$ is called the \emph{spectrum of
the decomposition. }Note that there are no restrictions on the number
$n$ in the definition of the spectrum, so if two spectra have different
length we will extend the shorter vector by concatenating zeros at
the end. We note that for a decomposition like (\ref{eq:SpecDecomp})
the trace is given by $\mathrm{tr}\left[x\right]=\sum_{i=1}^{n}\lambda_{i}$. 

Spectra are ordered by \emph{majorization}. Let $\lambda_{1}^{n}$
and $\mu_{1}^{n}$  be the spectra of two decompositions of the same
positive element.  Then $\lambda_{1}^{n}\succeq\mu_{1}^{n}$  if $\sum_{i=1}^{k}\lambda_{i}\geq\sum_{i=1}^{k}\mu_{i}$
for $k\leq n$. Note that in a general positive cone the majorization
ordering is a partial ordering. In special cases like the cone of
positive semidefinite matrices on a complex Hilbert space the decompositions
of the matrix form a lattice ordering with a unique maximal element,
but in general the set of decompositions may have several incompatible
maximal elements. 
\begin{defn}
Let $\mathcal{S}$ denote a convex set. Let $\sigma_{i},i=1,2,\dots,n$
states in the state space $S.$ Then $\left(\sigma_{i}\right)_{i}$
are said to be perfectly distinguishable if there exists tests $\phi_{i}$
such that $\phi_{i}\left(\sigma_{j}\right)=\delta_{ij}$. The states
$\sigma_{0}$ and $\sigma_{1}$ are said to be \emph{orthogonal} if
$\sigma_{0}$ and $\sigma_{1}$ are perfectly distinguishable in the
smallest face $F$ of $S$ that contain both $\sigma_{0}$ and $\sigma_{1}.$
If the the extreme points $\sigma_{1},\sigma_{2},\dots,\sigma_{n}$
of a decomposition are orthogonal then the decomposition is called
an \emph{orthogonal decomposition}.
\end{defn}
\begin{figure}[tbh]
\begin{centering}
\begin{tikzpicture}[line cap=round,line join=round,>=triangle 45,x=5.0cm,y=5.0cm] 
\draw[->,color=black] (0.,-0.16592592592592395) -- (0.,1.13185185185185); 
\foreach \y in {0,0.2,0.4,0.6,0.8,1} 
\draw[shift={(0,\y)},color=black] (2pt,0pt) -- (-2pt,0pt) node[left] {\footnotesize $\y$};
\clip(-0.17703703703703733,-0.16592592592592395) rectangle (1.363703703703703,1.13185185185185); 
\draw(0.8,0.5) circle (2.5cm); 
\draw [dash pattern=on 4pt off 4pt] (0.8,0.)-- (0.,0.); 
\draw (0.37,0.) -- (0.42,0.04); 
\draw (0.37,0.) -- (0.42,-0.04); 
\draw [dash pattern=on 4pt off 4pt] (0.8,1.)-- (0.,1.); 
\draw (0.37,1.) -- (0.42,1.04); 
\draw (0.37,1.) -- (0.42,0.96); 
\draw [dash pattern=on 2pt off 2pt] (0.8,0.4)-- (0.,0.4); 
\draw (0.37,0.4) -- (0.42,0.44); 
\draw (0.37,0.4) -- (0.42,0.36); 
\draw (0.8,1.)-- (0.8,0.); 
\draw (0.5,1) node[anchor=south] {$\phi$}; 
\draw (0.8,0) node[anchor=north] {$s_0$}; 
\draw (0.8,1) node[anchor=south] {$s_1$}; 
\draw (0.8,0.4) node[anchor= west] {$s$}; 
\begin{scriptsize} 
\draw [fill=black] (0.8,0.) circle (1.5pt); 
\draw [fill=black] (0.8,1.) circle (1.5pt); 
\draw [fill=black] (0.8,0.4) circle (1.5pt); 
\end{scriptsize} 
\end{tikzpicture}
\par\end{centering}
\caption{In the disc the points $\sigma_{0}$ and $\sigma_{1}$ are mutually
singular. The point $s$ has a unique decomposition into mutually
singular points because it is not the center of the disc.}
\end{figure}
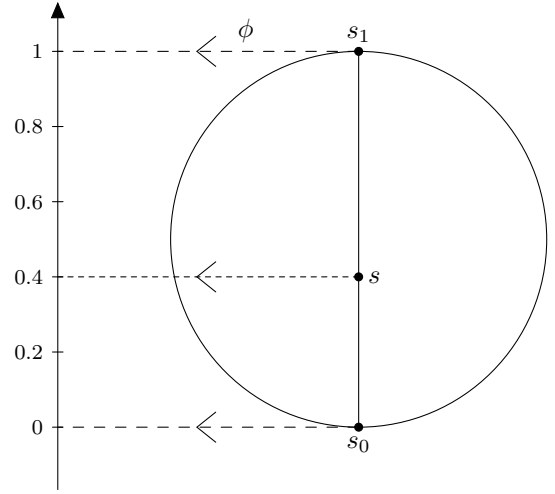

\begin{thm}
Let $\mathcal{S}$ denote a convex compact set of dimension $d$ and
let $x$ denote some element in the positive cone generated by $S$.
Then there exists an orthogonal decomposition of the form as in Equation
\ref{eq:SpecDecomp} such that $n\leq d+1$.
\end{thm}
\begin{IEEEproof}
See \cite[Thm. 2]{Harremoes2016d}.
\end{IEEEproof}
\begin{defn}
The\emph{ rank} of a convex set is the maximal number of orthogonal
states needed in an orthogonal decomposition of a state.
\end{defn}
\begin{example}
\label{exa:square}In the unit square with $\left(0,0\right),\left(1,0\right),\left(0,1\right)$
and $\left(1,1\right)$ as vertices the point, with coordinates $\left(\nicefrac{1}{2},\nicefrac{1}{4}\right)$
has an orthogonal decomposition with spectrum $\left(\nicefrac{1}{2},\nicefrac{1}{4},\nicefrac{1}{4}\right)$.
This spectrum majorizes the spectrum of any other decomposition of
this point, and it also majorizes the spectrum of any other point
in the square. The square has in total four points symmetrically arranged
with the same spectrum as $\left(\nicefrac{1}{2},\nicefrac{1}{4}\right)$.
\end{example}

\section{Spectral sets}

Any state in may be decomposed into orthogonal states, but such orthogonal
decompositions are only unique when the state space is a simplex.
Nevertheless there exists a type of convex sets where some weaker
type uniqueness holds.

\subsection{Spectrality conditions}
\begin{defn}[\cite{Barnum2015}]
An $n$-frame is a list of $n$ perfectly distinguishable pure states.
If any state $\sigma\in S$ has a decomposition $\sigma=\sum_{i}\lambda_{i}\cdot\sigma_{i}$
where $\sigma_{1},\sigma_{2,}\dots,\sigma_{n}$ is an $n$-frame then
the state space is said to satisfy weak spectrality.
\end{defn}
\begin{prop}
Let $S$ denote a state space such that 

1. If both $\rho\bot\sigma_{1}$ and $\rho\bot\sigma_{2}$ then $\rho\bot\frac{1}{2}\left(\sigma_{1}+\sigma_{2}\right)$.

2. If $\sigma_{1}\bot\sigma_{2}$ then $\sigma_{1}$ and $\sigma_{2}$
are perfectly distinguishable.

Then each state satisfies weak spectrality.
\end{prop}
\begin{defn}
If all orthogonal decompositions of a state have the same spectrum
then the common spectrum is called it the \emph{spectrum of the state}
and the state is said to be \emph{spectral}. We say that a state space
$\mathcal{S}$ is \emph{spectral} if all states in $\mathcal{S}$
are spectral. 
\end{defn}
Often we are interested in a stronger condition.
\begin{defn}
Let $\sigma$ denote a state with the orthogonal decompositions $\sigma=\sum s_{i}\sigma_{i}=\sum r_{j}\rho_{j}.$
We say that the decompositions are strictly spectral if $\sum f\left(s_{i}\right)\sigma_{i}=\sum f\left(r_{j}\right)\rho_{j}$
for any real valued function $f$. A state is strictly spectral if
any orthogonal decomposition is strictly spectral. A state space is
strictly sprectral if any state is strictly spectral.
\end{defn}
The important case is when the function $f$ equals $1_{\lambda}$
for which we get that $\sum s_{i}\sigma_{i}=\sum r_{j}\rho_{j}$ implies
that $\sum_{s_{i}=\lambda}\sigma_{i}=\sum_{r_{j}=\lambda}\rho_{j}.$
An element of the form $\sum_{s_{i}=\lambda}\sigma_{i}$ is called
an idempotent. Strict spectrality means that a state has a unique
decomposition $\sigma=\sum\lambda_{k}e_{k}$ where $\lambda_{k}$are
distinct eigenvalues and $e_{k}$ are orthogonal idempotents. We note
that if the state space is strictly spectral then for any element
$x$ in the generated real vector space we have $y\geq0$ if and only
if there exists $x$ such that $y=x^{2}.$ In particular $\sum x_{i}^{2}=0$
if and only if $x_{i}=0$ for all $i.$ 

\subsection{Jordan algebras}

The notion of a spectral set is related to self-duality of the cone
of positive elements, which leads to the following conjecture. 
\begin{conjecture}
If a finite dimensional convex compact set is spectral and has a transitive
symmetry group then the convex set can be represented as positive
elements with trace 1 in a simple Jordan algebra .
\end{conjecture}
The density matrices with complex entries play a crusial role in the
mathematical theory of quantum mechanics and it is well-known that
the density matrices is a spectral set. For each density matrix the
spectrum equals the usual spectrum calculated as roots of the characteristic
polynomium. In the 1930'ties P. Jordan generalized the notion of Hermitean
complex matrix to the notion of a Jordan algebra in an attempt to
provide an alternative to the complex Hilbert spaces as the mathematical
basis of quantum mechanics. For instance the complex Hermitean matrices
form a Jordan algebra with the \emph{quasi-product} defined by 
\begin{equation}
x\circ y=\frac{1}{2}\left(\left(x+y\right)^{2}-x^{2}-y^{2}\right).\label{eq:quasimult}
\end{equation}
A direct expansion of the squares lead to $x\circ y=\frac{1}{2}\left(xy+yx\right).$ 

An formally real Jordan algebra is an algebra with composition $\circ$
that is commutative and satifies the Jordan identity  
\[
\left(x\circ y\right)\circ\left(x\circ x\right)=x\circ\left(y\circ\left(x\circ x\right)\right).
\]
 Further it is assumed that 
\[
\sum_{i=1}^{n}x_{i}^{2}=0
\]
 implies that $x_{i}=0$ for all $i.$ In an formally real Jordan
algebra we write $x\geq0$ if $x$ is a sum of squares.

For a matrix $\left(M_{mn}\right)$ the trace $\mathrm{tr}$ is defined
by $\mathrm{tr}\left[M\right]=\sum_{n}M_{nn}.$ The trace vanish on
commutators and associators. For an associative algebra it means that
$\mathrm{tr}\left[MN\right]=\mathrm{tr}\left[NM\right]$ and for a
Jordan algebra it means that $\mathrm{tr}\left[\left(a\circ b\right)\circ c\right]=\mathrm{tr}\left[a\circ\left(b\circ c\right)\right]$
(see \cite{Gordon1974} for details). For a Jordan algebra one can
define an inner product by $\left\langle x,y\right\rangle =\mathrm{tr}\left[x\circ y\right].$
With this inner product the positive cone becomes self-dual.

In any finite dimensional formally real Jordan algebra we may define
the density operators as the positive elements with trace 1. Then
the density operators of a Jordan algebra is a spectral set. Any formally
real Jordan algebra is a sum of simple formally real Jordan algebras.
There are 5 types of simple Eucledian Jordan algebras leading to the
following convex sets:

\textbf{Spin factor} A unit ball in $d$ real dimensions.

\textbf{Real} $n\times n$ density matrices over the real numbers.

\textbf{Complex} $n\times n$ density matrices over complex numbers

\textbf{Quaternionic} $n\times n$ density matrices over quaternionians.

\textbf{Albert} $3\times3$ density matrices with octonian entries. 

The first four types are called the \emph{special types} and the last
one is called the \emph{exceptional type}. In each of the four special
Jordan algebras the product is defined by (\ref{eq:quasimult}) from
an associative product. This is not possible for the Albert algebra,
which is the reason that it is said to be exceptional. See \cite{McCrimmon2004}
for general results on Jordan algebras and \cite{Adler1995} for details
about quantum mechanics based on quatonians. The $2\times2$ matrices
have real, complex, quaternionic or octonionic entries can be identified
with spin factors with $d=2,$ $d=3,$ $d=5,$ and $d=9.$ The most
important example of a spin factor is the qubit.

A sum of different simple Jordan algebras does not fulfill the strong
symmetry mentioned in \cite{Barnum2015} although it is spectral and
fulfill projectivity. This was left as an open question in \cite{Barnum2015}.

\subsection{Separable states}

Let $\mathbb{H}_{1}\otimes\mathbb{H}_{2}$ denote a tensor product
of two complex Hilbert spaces. Then $\mathbb{B}\left(\mathbb{H}_{1}\otimes\mathbb{H}_{2}\right)=\mathbb{B}\left(\mathbb{H}_{1}\right)\otimes\mathbb{B}\left(\mathbb{H}_{2}\right).$
The separable states are mixtures of states of the form $\sigma_{1}\otimes\sigma_{2}$
where $\sigma_{i}$ is a density operator in $\mathbb{B}\left(\mathbb{H}_{i}\right).$
We will show that the set of separable states is spectral. 

First we note that for $A\in\mathbb{B}\left(\mathbb{H}_{1}\otimes\mathbb{H}_{2}\right)$
and $B\in\mathbb{B}\left(\mathbb{H}_{1}\right)$ and $C\in\mathbb{B}\left(\mathbb{H}_{1}\right)$we
have 
\[
\mathrm{tr}\left(A\left(B\otimes C\right)\right)=\mathrm{tr}_{1}\left(\mathrm{tr}_{2}\left(A\right)B\right)\cdot tr_{2}\left(\mathrm{tr}_{1}\left(A\right)C\right)
\]
where $\mathrm{tr}_{1}$ and $\mathrm{tr}_{2}$ denote partial traces.
Therefore $\mathrm{tr}\left(A\left(B\otimes C\right)\right)$ is extreme
when $\mathrm{tr}_{1}\left(\mathrm{tr}_{2}\left(A\right)B\right)$
and $\mathrm{tr}_{2}\left(\mathrm{tr}_{1}\left(A\right)C\right)$
are extreme. Now $\mathrm{tr}_{1}\left(\mathrm{tr}_{2}\left(A\right)B\right)$
is minimal if $B$ is supporten on the eigen space of the minimal
eigenvalue of $\mathrm{tr}_{2}\left(A\right)$ and $\mathrm{tr}_{1}\left(\mathrm{tr}_{2}\left(A\right)B\right)$
is maximal if $B$ is supporten on the eigen space of the maximal
eigenvalue of $\mathrm{tr}_{2}\left(A\right).$ In particular $B_{1}\otimes C_{1}$
and $B_{2}\otimes C_{2}$ are orthogonal as elements of the set of
separable states if $B_{1}\bot B_{2}$or $C_{1}\bot C_{2}.$ Since
$\mathrm{tr}\left(\left(B_{1}\otimes C_{1}\right)\left(B_{2}\otimes C_{2}\right)\right)=\mathrm{tr}\left(B_{1}B_{2}\right)\mathrm{tr}\left(C_{1}C_{2}\right)$
we have that $B_{1}\otimes C_{1}$ and $B_{2}\otimes C_{2}$ are orthogonal
as elements of the set of separable states if and only if $B_{1}\otimes C_{1}$
and $B_{2}\otimes C_{2}$ are orthogonal under the Hilbert-Schmidt
inner product on $\mathbb{B}\left(\mathbb{H}_{1}\otimes\mathbb{H}_{2}\right).$

Now any separable state can be decomposed into $\sum_{i}\lambda_{i}B_{i}\otimes C_{i}.$
where $B_{i}\otimes C_{i}$ are orthogonal pure states on $\mathbb{B}\left(\mathbb{H}_{1}\otimes\mathbb{H}_{2}\right)$
and since the density matrices on $\mathbb{B}\left(\mathbb{H}_{1}\otimes\mathbb{H}_{2}\right)$
form a spectral set the coefficients $\lambda_{i}$ is uniquely determined.
The separable states satisfy Bell type inequality and each Bell inequality
determines a face of the set of separable states. These faces do not
have complementary faces so the set of separable states does not satisfy
the property called projectivity.

\subsection{Symmetry}

Recal that a convex set $C$ is balanced about the origin if $P\in C$
implies $-P\in C$ where $-P$ denotes the point with opposite sign
of all coordinates. A spectral set of rank 2 is balanced, i.e. it
is symmetric around a central point and all boundary points are extreme.
Two states in the set are orthogonal if and only if they are antipodal.
Any state can be decomposed into two antipodal states. If the state
is not the center of the balanced set this is the only orthogonal
decomposition. The center can be decomposed into a  $\nicefrac{1}{2}$
and $\nicefrac{1}{2}$ mixture of any pair of antipodal points.
\begin{prop}
In two dimensions a simplex and a balanced set are the only types
of spectral sets.
\end{prop}
A state space of rank 2 is said to have \emph{symmetric transission
probabilities} \cite[Def. 9.2 (iii)]{Alfsen2003} if for any states
$\sigma_{1}$ and $\sigma_{2}$there exists test $\phi_{i}$ such
that $\phi_{i}\left(\sigma_{i}\right)=1$ and $\min_{\sigma}\phi_{i}\left(\sigma\right)=0.$
\begin{thm}
A spectral state space of rank 2 with symmetric transmission probabilities
can be represented by a spin factor.
\end{thm}
\begin{IEEEproof}
Let $C$ denote an intersection between the state space and a two
dimensional hyperplane through the center $c$ of the state space.
Let $\sigma_{1}$and $\sigma_{2}$ denote states such that $\phi_{1}\left(\sigma_{2}\right)=\phi_{2}\left(\sigma_{1}\right)=\nicefrac{1}{2}.$
Define $\tilde{\phi}_{i}=2\cdot\phi_{i}-1$ so that $\tilde{\phi}_{i}\left(\sigma_{i}\right)=1$
and $\tilde{\phi}_{i}\left(c\right)=0$. For any state $\sigma$ we
may define the coordinates by $x=\tilde{\phi}_{1}\left(\sigma\right)$
and $y=\tilde{\phi}_{2}\left(\sigma\right).$ Let $\phi$denote a
test that equals 1 on $\sigma$and equals 0 on the antipodal of $\sigma.$
Let $\tilde{\phi}=2\cdot\phi-1.$ If the boundary of the state space
at $\sigma$ only has one supporting hyperplane then $\tilde{\phi}$is
uniquely determined. Then $\tilde{\phi}\left(\sigma_{1}\right)=x$
and $\tilde{\phi}\left(\sigma_{2}\right)=y.$ Therefore the projection
of $\sigma_{1}$along the supporting hyperplan has coordinates $\left(x^{2},xy\right)$
and the projection of $\sigma_{2}$ has coordinates $\left(xy,y^{2}\right).$
Therefore the vectors $\left(x^{2}-1,xy\right)$ and $\left(xy,y^{2}-1\right)$
are parallel so that 
\begin{align*}
\left|\begin{array}{cc}
x^{2}-1 & xy\\
xy & y^{2}-1
\end{array}\right| & =0\\
1-x^{2}-y^{2} & =0
\end{align*}
so that $\sigma$ has coordinates that lie on a unit circle. Almost
all points on the boundary of $C$ has a unique supporting hyperplane.
Therefore almost all points lie on a circle so all points must lie
on a circle.
\end{IEEEproof}
A state space is said to be \emph{symmetric} if any $n$-frame can
be transformed into any other $n$-frame by an affine map of the state
space into itself. 
\begin{conjecture}
A symmetric spectral state space is either a simplex or it can be
represented as density elements of a formally real Jordan algebra.
\end{conjecture}

\section{Quantization of Energy}

Often the term quantum theory is used only for systems that have physical
implementations. Here we will justify that the term ``quantum''
is used for Jordan algebras and even more general cases. Assume that
$S$ is finite dimensional state space. Let $g_{t}:S\to S$ denote
the transformation that maps a state at time 0 into state at time
$t$. We assume that $g_{s+t}=g_{s}\circ g_{t}$ so that we get a
representation of $\mathbb{R}_{0,+}.$ We shall restrict the discussion
to the harmonic oscillator. In classical physics a harmonic oscilator
has a period $T$ such that the state at time $t$ equals the state
at time $t+T.$ Therefore we may call a system with state space $S$
a harmonic oscillator if $g_{t+T}=g_{t}$ for all $t\in\mathbb{R}_{0,+}$
and this representation can be extended to $\mathbb{R}.$ Since the
representation is periodic it may be considered as a representation
of $\mathbb{T}=\mathbb{R}/\tau$ where $\tau$ is the circle constant
$2\pi.$ We say that the representation $t\to g_{t}$ of $\mathbb{T}$
into transformations $g_{t}:S\to S$ is irreducible if any convex
subset $S'\subseteq S$ that is is invariant under $g_{t}$ spans
$S.$ 
\begin{prop}
If $t\to g_{t}$ is an irreducible representation of $\mathbb{T}$
on the state space $S$ then $S$ is a point or $S$ has the shape
of a disk. If $S$ is a disk then there exists a $n\in\mathbb{N}$
such that $t$ is mapped into a rotation by the angle $n\cdot t.$ 
\end{prop}
In the standard Hilbert space formalism the representation characterized
by $n$ is described by a Hamiltonian with energy $\hbar\omega\left(n+1/2\right).$
If $\mathbb{T}$ is represented on a finite dimensional state space
but the representation is not irreducible then the representation
can be decomposed into irreducible representations each characterized
by a specific energy. In this sense energy is quantized. In the standard
formulation of quantum mechanics the energy is represented by a operator
that is an observable, but for other Jordan algebras there exists
connected symmetry groups that cannot be represented by observables
\cite{Barnum2014}.

\section{Optimization}

Let $\mathcal{A}$ denote a subset of the feasible measurements such
that $a\in\mathcal{A}$ maps the convex set $\mathcal{S}$ into a
distribution on the real numbers i.e. the distribution of a random
variable. The elements of $\mathcal{A}$ may represent feasible \emph{actions}
(decisions) that lead to a payoff like the score of a statistical
decision, the energy extracted by a certain interaction with the system,
(minus) the length of a codeword of the next encoded input letter
using a specific code book, or the revenue of using a certain portfolio.
For each $\sigma\in\mathcal{S}$ we define 
\[
\left\langle a,\sigma\right\rangle =E\left[a\left(\sigma\right)\right].
\]
 and 
\[
F\left(\sigma\right)=\sup_{a\in\mathcal{A}}\left\langle a,\sigma\right\rangle .
\]
 Without loss of generality we may assume that the set of actions
$\mathcal{A}$ is closed so that we may assume that there exists $a\in\mathcal{A}$
such that $F\left(\sigma\right)=\left\langle a,\sigma\right\rangle $
and in this case we say that $a$ is optimal for $\sigma.$ We note
that $F$ is convex but $F$ need not be strictly convex.

If $F\left(\sigma\right)$ is finite then we define \emph{the regret}
of the action $a$ by 
\[
D_{F}\left(\sigma,a\right)=F\left(\sigma\right)-\left\langle a,\sigma\right\rangle .
\]
 
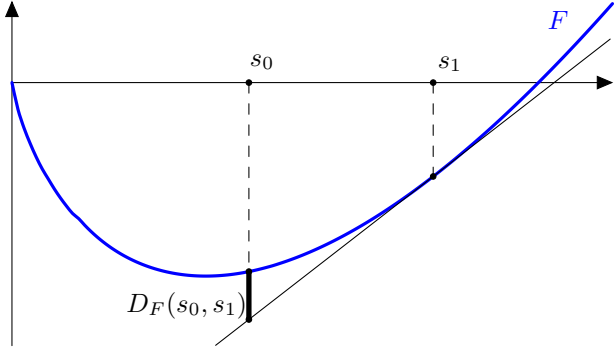
\begin{figure}[tbh]
\begin{centering}
\begin{tikzpicture}[scale=0.70,line cap=round,line join=round,>=triangle 45,x=10.0cm,y=10.0cm] 
\draw[->,color=black] (0.,0.) -- (1.1419478737997253,0.); 
\draw[->,color=black] (0.,-0.5) -- (0.,0.15580246913580265); 
\clip(-0.06431,-0.5008779149519884) rectangle (1.1419478737997253,0.15580246913580265); 
\draw[line width=1.2pt,color=blue,smooth,samples=100,domain=0:1.1419478737997253] plot(\x,{(\x)*ln((\x)+1.0E-4)}); 
\draw (0.3868, -0.4994) -- (1.1362, 0.0827); \draw [dash pattern=on 4pt off 4pt] (0.8,0.)-- (0.8,-0.178414847300847); 
\draw [line width=2.pt] (0.45,-0.450314607074793)-- (0.45,-0.35922847440746253); 
\draw [dash pattern=on 4pt off 4pt] (0.45,0.)-- (0.45,-0.35922847440746253); 
\draw (0.4343468211923174,0.07152263374485619) node[anchor=north west] {$s_0$}; 
\draw (0.790781098312178,0.0680109739369001) node[anchor=north west] {$s_1$}; 
\draw (0.2,-0.38) node[anchor=north west] {$D_{F}(s_0 ,s_1)$}; 
\draw (1,0.1540466392318246) node[anchor=north west, color=blue] {$F$}; 
\draw [fill=black] (0.8,-0.178414847300847) circle (1.5pt); 
\draw [fill=black] (0.45,-0.35922847440746253) circle (1.5pt); 
\draw [fill=black] (0.45,-0.450314607074793) circle (1.5pt); 
\draw [fill=black] (0.45,0.) circle (1.5pt); 
\draw [fill=black] (0.8,0.) circle (1.5pt); 
\end{tikzpicture}
\par\end{centering}
\caption{The regret equals the vertical between curve and tangent.\label{breg}}
\end{figure}

If $a_{i}$ are actions and $\left(t_{i}\right)$ is a probability
vector then we we may define the mixed action $\sum t_{i}\cdot a_{i}$
as the action where we do $a_{i}$ with probability $t_{i}.$ We note
that $\left\langle \sum t_{i}\cdot a_{i},\sigma\right\rangle =\sum t_{i}\cdot\left\langle a_{i},\sigma\right\rangle .$
We will assume that all such mixtures of feasible actions are also
feasible. If $a_{1}\left(\sigma\right)\geq a_{2}\left(\sigma\right)$
almost surely for all states we say that $a_{1}$ dominates $a_{2}$
and if $a_{1}\left(\sigma\right)>a_{2}\left(\sigma\right)$ almost
surely for all states $\sigma$ we say that $a_{1}$ strichtly dominates
$a_{2}.$ All actions that are dominated may be removed from $\mathcal{A}$
without changing the function $F.$ Let $A_{F}$ denote the set of
measurements $m$ such that $\left\langle m,\sigma\right\rangle \leq F\left(\sigma\right).$
Then $F\left(\sigma\right)=\sup_{a\in\mathcal{A}_{F}}\left\langle a,\sigma\right\rangle .$
Therefore we may replace $\mathcal{A}$ by $\mathcal{A}_{F}$ without
changing the optimization problem.
\begin{defn}
If $F\left(\sigma\right)$ is finite \emph{the regret} of the action
$a$ is defined by 
\begin{equation}
D_{F}\left(\sigma,a\right)=F\left(\sigma\right)-\left\langle a,\sigma\right\rangle 
\end{equation}
\end{defn}
\begin{prop}
The regret $D_{F}$ has the following properties:
\end{prop}
\begin{itemize}
\item $D_{F}\left(\sigma,a\right)\geq0$ with equality if $a$ is optimal
for $\sigma$.
\item $\sigma\to D_{F}\left(\sigma,a\right)$ is a convex function.
\item If $\bar{a}$ is optimal for the state $\bar{\sigma}=\sum t_{i}\cdot\sigma_{i}$
where $\left(t_{1},t_{2},\dots,t_{\ell}\right)$ is a probability
vector then 
\[
\sum t_{i}\cdot D_{F}\left(\sigma_{i},a\right)=\sum t_{i}\cdot D_{F}\left(\sigma_{i},\bar{a}\right)+D_{F}\left(\bar{\sigma},a\right).
\]
 
\item $\sum t_{i}\cdot D_{F}\left(\sigma_{i},a\right)$ is minimal if $a$
is optimal for $\bar{\sigma}=\sum t_{i}\cdot\sigma_{i}$.
\end{itemize}
If the state is not know exactly but we know that $\sigma$ is one
of the states $\sigma_{1},\sigma_{2},\dots,\sigma_{n}$ then the \emph{minimax
regret} is defined as 
\[
C_{F}=\inf_{a}\sup_{i}D_{F}\left(\sigma_{i},a\right).
\]
We have the following result.
\begin{thm}
For any set of actions 
\[
C_{F}=\sup_{\vec{t}}\inf_{a}\sum_{i}t_{i}\cdot D_{F}\left(\sigma_{i},a\right)
\]
where the supremum is taken over all probability vectors $\vec{t}$
supported on $\mathcal{S}$. 
\end{thm}
This result can improved.
\begin{thm}
If $\left(t_{1},t_{2},\dots,t_{n}\right)$ is a probability vector
on the states $\sigma_{1},\sigma_{2},\dots,\sigma_{n}$ with $\bar{\sigma}=\sum t_{i}\cdot\sigma_{i}$
and $a_{opt}$ is the optimal action for minimax regret then 
\[
C_{F}\geq\inf_{a}\sum t_{i}\cdot D_{F}\left(\sigma_{i},a\right)+D_{F}\left(\bar{\sigma},a_{opt}\right).
\]
If $a$ is an action and $\sigma_{opt}$ is optimal then 
\[
\sup_{i}D_{F}\left(\sigma_{i},a\right)\geq C_{F}+D_{F}\left(\sigma_{opt},a\right).
\]
\end{thm}
\begin{IEEEproof}
See \cite[Thm. 2]{Harremoes2017}.
\end{IEEEproof}

\subsection{Bregman divergences}

If the state is $\rho$ but one acts as if the state were $\sigma$
one suffers a regret that equals the difference between what one achieves
and what could have been achieved. 
\begin{defn}
\label{def:regret} If $F\left(\rho\right)$ is finite then we define
\emph{the regret of the state} $\sigma$ as 
\[
D_{F}\left(\rho,\sigma\right)=\inf_{a}D_{F}\left(\rho,a\right)
\]
where the infimum is taken over actions $a$ that are optimal for
$\sigma.$ 
\end{defn}
If the state $\sigma$ has the unique optimal action $a$ then 
\begin{equation}
F\left(\rho\right)=D_{F}\left(\rho,\sigma\right)+\left\langle a,\rho\right\rangle 
\end{equation}
so the function $F$ can be reconstructed from $D_{F}$ except for
an affine function of $\rho.$ The closure of the convex hull of the
set of functions $\sigma\to\left\langle a,\sigma\right\rangle $ is
uniquely determined by the convex function $F.$ 

The regret is called a \emph{Bregman divergence} if it can be written
in the following form
\begin{align}
D_{F}\left(\rho,\sigma\right) & =F\left(\rho\right)-\left(F\left(\sigma\right)+\left\langle \rho-\sigma,\nabla F\left(\sigma\right)\right\rangle \right)
\end{align}
where $\left\langle \cdot,\cdot\right\rangle $ denotes some inner
product. In the context of forecasting and statistical scoring rules
the use of Bregman divergences dates back to \cite{Hendrickson1971}. 
\begin{thm}
The following conditions are equivalent.

For each state $s$ and all actions $a_{1}$ and $a_{2}$ such that
$F\left(\sigma\right)=\left\langle a_{1},\sigma\right\rangle =\left\langle a_{2},\sigma\right\rangle $
we have $E\circ a_{1}=E\circ a_{2}.$

The function $F$ is differentiable.

The regret $D_{F}$ is a Bregman divergence.
\end{thm}
We note that if $D_{F}$ is a Bregman divergence and $\sigma$ minimizes
$F$ then $\nabla F\left(\sigma\right)=0$ so that the formula for
the Bregman divergence reduces to 
\[
D_{F}\left(\rho,\sigma\right)=F\left(\rho\right)-F\left(\sigma\right).
\]
Bregman divergences satisfy the \emph{Bregman identity} 
\begin{equation}
\sum t_{i}\cdot D_{F}\left(\rho_{i},\sigma\right)=\sum t_{i}\cdot D_{F}\left(\rho_{i},\bar{\rho}\right)+D_{F}\left(\bar{\rho},\sigma\right),\label{eq:Bregmanid}
\end{equation}
but if $F$ is not differentiable this identity can be violated.
\begin{example}
Let the state space be the interval $\left[0,1\right]$ with two actions
$a_{0}\left(\sigma\right)=1-2\sigma$ and $a_{1}\left(\sigma\right)=2\sigma-1.$
Let $\sigma_{0}=0$ and $\sigma_{1}=1.$ Let further $t_{0}=\nicefrac{1}{3}$
and $t_{1}=\nicefrac{2}{3}.$ Then $\bar{\sigma}=\nicefrac{2}{3}.$
If $\sigma=\nicefrac{1}{2}$ then 
\[
\sum t_{i}\cdot D_{F}\left(\sigma_{i},s\right)=0
\]
but 
\begin{multline*}
\sum t_{i}\cdot D_{F}\left(\sigma_{i},\bar{\sigma}\right)=\\
\frac{1}{3}\cdot\left(a_{0}\left(0\right)-a_{1}\left(0\right)\right)+\frac{2}{3}\cdot\left(a_{1}\left(1\right)-a_{1}\left(1\right)\right)\\
=\frac{1}{3}\cdot\left(1-\left(-1\right)\right)=\frac{2}{3}.
\end{multline*}
 We also have $D_{F}\left(\bar{\sigma},\sigma\right)=0.$ Clearly
the Bregman identity (\ref{eq:Bregmanid}) is violated.

Assume that $\bar{a}$ is optimal for $\bar{\rho}.$ Then 
\begin{align*}
\sum t_{i}\cdot\left(F\left(\rho_{i}\right)\right) & =\sum t_{i}\cdot\left(D_{F}\left(\rho_{i},\bar{\rho}\right)+\left\langle \bar{a},\rho_{i}\right\rangle \right)\\
 & =\sum t_{i}\cdot D_{F}\left(\rho_{i},\bar{\rho}\right)+\left\langle \bar{a},\sum t_{i}\cdot\rho_{i}\right\rangle \\
 & =\sum t_{i}\cdot D_{F}\left(\rho_{i},\bar{\rho}\right)+\left\langle \bar{a},\bar{\rho}\right\rangle \\
 & =\sum t_{i}\cdot D_{F}\left(\rho_{i},\bar{\rho}\right)+F\left(\bar{\rho}\right).
\end{align*}
Therefore
\[
\sum t_{i}\cdot D_{F}\left(\rho_{i},\bar{\rho}\right)=\sum t_{i}\cdot\left(F\left(\rho_{i}\right)\right)-F\left(\bar{\rho}\right).
\]
\end{example}

\section{Sufficiency conditions}

In this section we will introduce various conditions on a Bregman
divergence. Under some mild conditions they turn out to be equivalent.
\begin{thm}
Let $D_{F}$ denote a regret function defined on an arbitrary state
space. Then we have the implications $3.\Rightarrow4.\Rightarrow5.\Rightarrow6.\Rightarrow1.\Rightarrow2.$

1. The function $F$ equals entropy times a constant plus an affine
function.

2. The regret $D_{F}$ is proportional to information divergence.

3. The regret function satisfies strong sufficiency.

4. The regret is monotone, i.e. it satisfies the data processing inequality.

5. The regret is sufficiency stable.

6. The regret is local.

The conditions 1., 2., and 6. are equivalent on formally real Jordan
algebras with at least three orthogonal states.. 
\end{thm}

\subsection{Entropy in Jordan algebras}
\begin{defn}
Let $x$ denote an element in a positive cone. The \emph{entropy}
of $x$ is be defined as 
\[
H\left(x\right)=\inf\left(-\sum_{i=1}^{n}\lambda_{i}\ln\left(\lambda_{i}\right)\right)
\]
where the infimum is taken over all spectra of $x$. 
\end{defn}
Since entropy is decreasing under majorization the entropy of $x$
is attained at an orthogonal decomposition. This definition extends
a similar definition of the entropy of a state as defined by Uhlmann
\cite{Uhlmann1970}. 

In general this definition of entropy does not provide a concave function
on the positive cone. For instance the entropy of points in the square
from Example \ref{exa:square} has local maximum in the four points
with maximal spectrum in the majorization ordering. A characterization
of the convex sets with concave entropy functions is lacking.
\begin{prop}
Assume that the entropy function $H$ on a state space is strictly
concave. Then any uniform mixture of orthogonal pure states give the
maximum entropy state.
\end{prop}
\begin{IEEEproof}
Assume that $\sigma=\frac{1}{n}\sum_{i=1}^{n}\sigma_{i}$ and $\rho=\frac{1}{n}\sum_{i=1}^{n}\rho_{i}$
are decompositions into orthogonal pure states. Then $H\left(\sigma\right)=H\left(\rho\right)=\ln\left(n\right)$
and $H\left(\frac{\sigma+\rho}{2}\right)\geq\ln\left(n\right)$with
equality, which implies that $\sigma=\rho$
\end{IEEEproof}
The entropy is defined as for general convex set and we will prove
that $H$ is a concave on the cone of positive elements in a Jordan
algebra. The following exposition is inspired by similar result for
complex matrix algebras stated in \cite{Petz2008}, but the proofs
have been changed so that they are valid on Jordan algebras.
\begin{lem}
For elements $A$ and $B$ in a Jordan algebra and any analytic function
$f$ we have 
\[
\left.\frac{\mathrm{d}}{\mathrm{d}t}\mathrm{tr}\left[f\left(A+tB\right)\right]\right\vert _{t=0}=\mathrm{tr}\left[f^{\prime}\left(A\right)\circ B\right].
\]
\end{lem}
\begin{IEEEproof}
First assume that $f\left(z\right)=z^{r}.$ Then 
\begin{align*}
\left.\frac{\mathrm{d}}{\mathrm{d}t}\mathrm{tr}\left[f\left(A+tB\right)\right]\right\vert _{t=0} & =\mathrm{tr}\left[\left.\frac{\mathrm{d}}{\mathrm{d}t}\left(A+tB\right)^{r}\right\vert _{t=0}\right]\\
 & =\mathrm{tr}\left[\sum_{i=0}^{r-1}A^{i}\circ B\circ A^{r-1-i}\right]\\
 & =\mathrm{tr}\left[\sum_{i=0}^{r-1}A^{i}\circ A^{r-1-i}\circ B\right]\\
 & =\mathrm{tr}\left[n\cdot A^{r-1}\circ B\right]\\
 & =\mathrm{tr}\left[f^{\prime}\left(A\right)\circ B\right].
\end{align*}
As a consequence the theorem holds for any polynomial and also for
any analytic function because such functions can be approximated by
polynomials. 
\end{IEEEproof}
\begin{lem}
In a real Jordan algebra the following formula holds for any analytic
function $f$
\[
\left.\frac{\mathrm{d}^{2}}{\mathrm{d}t^{2}}\mathrm{tr}\left[f\left(A+tB\right)\right]\right\vert _{t=0}=\sum_{k,\ell}a_{k,l}\mathrm{tr}\left[\left(E_{k}\circ B\right)\circ\left(E_{\ell}\circ B\right)\right]
\]
where $A=\sum_{k}\lambda_{k}E_{k}$ is an orthogonal decomposition
and 
\[
a_{k,\ell}=\begin{cases}
\frac{f^{\prime}\left(\lambda_{k}\right)-f^{\prime}\left(\lambda_{\ell}\right)}{\lambda_{k}-\lambda_{\ell}} & \text{for}\,\lambda_{k}\neq\lambda_{\ell},\\
f^{\prime\prime}\left(\lambda_{k}\right) & \text{for}\,\lambda_{k}=\lambda_{\ell}.
\end{cases}
\]
\end{lem}
\begin{IEEEproof}
Let $f$ denote the function $z^{r}.$ Then
\begin{multline*}
\left.\frac{\mathrm{d}^{2}}{\mathrm{d}t^{2}}\mathrm{tr}\left[f\left(A+tB\right)\right]\right\vert _{t=0}=\left.\frac{\mathrm{d}}{\mathrm{d}t}\mathrm{tr}\left[f'\left(A+tB\right)\circ B\right]\right\vert _{t=0}\\
=\left.\frac{\mathrm{d}}{\mathrm{d}t}\mathrm{tr}\left[r\cdot\left(A+tB\right)^{r-1}\circ B\right]\right\vert _{t=0}\\
=r\cdot\mathrm{tr}\left[\sum_{j=0}^{r-2}\left((((A^{j}\circ B)\stackrel{r-j-2\,\textrm{times}}{\overbrace{\circ A)\circ A)\circ\dots\circ A}}\right)\circ B\right]\\
=r\cdot\mathrm{tr}\left[\sum_{j=0}^{r-2}\left(A^{j}\circ B\right)\circ\left(A^{r-2-j}\circ B\right)\right].
\end{multline*}
Assume $A=\sum_{k}\lambda_{k}E_{k}$ where $E_{k}$ are orthogonal
pure states. Then
\begin{multline*}
\left.\frac{\mathrm{d}^{2}}{\mathrm{d}t^{2}}\mathrm{tr}\left[f\left(A+tB\right)\right]\right\vert _{t=0}=\\
r\cdot\mathrm{tr}\left[\sum_{j=0}^{r-2}\left(\sum_{k}\lambda_{k}^{j}\cdot E_{k}\circ B\right)\circ\left(\sum_{\ell}\lambda_{\ell}^{r-2-j}\cdot E_{\ell}\circ B\right)\right]\\
=r\cdot\sum_{j=0}^{r-2}\mathrm{\sum_{k,\ell}\lambda_{k}^{j}\lambda_{\ell}^{r-2-j}tr}\left[\left(E_{k}\circ B\right)\circ\left(E_{\ell}\circ B\right)\right]\\
=r\cdot\mathrm{\sum_{k,\ell}\left(\sum_{j=0}^{r-2}\lambda_{k}^{j}\lambda_{\ell}^{r-2-j}\right)tr}\left[\left(E_{k}\circ B\right)\circ\left(E_{\ell}\circ B\right)\right]\\
=\mathrm{\sum_{k,\ell}\left(r\cdot\sum_{j=0}^{r-2}\lambda_{k}^{j}\lambda_{\ell}^{r-2-j}\right)tr}\left[\left(E_{k}\circ B\right)\circ\left(E_{\ell}\circ B\right)\right].
\end{multline*}
Now 
\begin{align*}
\sum_{j=0}^{r-2}\lambda_{k}^{j}\lambda_{\ell}^{r-2-j} & =\begin{cases}
\frac{r\cdot\lambda_{k}^{r-1}-r\cdot\lambda_{\ell}^{r-1}}{\lambda_{k}-\lambda_{\ell}} & \text{for}\,\lambda_{k}\neq\lambda_{\ell},\\
r\cdot\left(r-1\right)\lambda_{k}^{r-2} & \text{for}\,\lambda_{k}=\lambda_{\ell}
\end{cases}\\
 & =\begin{cases}
\frac{f^{\prime}\left(\lambda_{k}\right)-f^{\prime}\left(\lambda_{\ell}\right)}{\lambda_{k}-\lambda_{\ell}} & \text{for}\,\lambda_{k}\neq\lambda_{\ell},\\
f^{\prime\prime}\left(\lambda_{k}\right) & \text{for}\,\lambda_{k}=\lambda_{\ell}.
\end{cases}
\end{align*}
Since the formula holds of all powers, it also holds for all polynomials
and for all analytic functions because these can be approximated by
polynomials.
\end{IEEEproof}
\begin{thm}
In a formally real Jordan algebra the entropy function is a concave
function on the positive cone.
\end{thm}
\begin{IEEEproof}
Let $f$ denote the holomorphic function $f\left(z\right)=-z\ln z,~z>0.$
We have to prove that $\mathrm{tr}\left[f\left(\left(1-t\right)A+tX\right)\right]=\mathrm{tr}\left[f\left(A+tB\right)\right]$
is concave where $B=X-A$.  We have 
\[
\left.\frac{\mathrm{d}^{2}}{\mathrm{d}t^{2}}\mathrm{tr}\left[f\left(A+tB\right)\right]\right\vert _{t=0}=\sum_{k,\ell}a_{k,l}\mathrm{tr}\left[\left(E_{k}\circ B\right)\circ\left(E_{\ell}\circ B\right)\right]
\]
and the coefficients $a_{k,\ell}$ are negative because $f$ is concave.
We need to prove that $\mathrm{tr}\left[\left(E_{k}\circ B\right)\circ\left(E_{\ell}\circ B\right)\right]\geq0,$
but a formally real Jordan algebra can be written as a sum of simple
Jordan algebras so it is sufficient to prove positivity on simple
algebras.

Assume that the Jordan algebra is special we have
\begin{multline*}
\mathrm{tr}\left[\left(E_{k}\circ B\right)\circ\left(E_{\ell}\circ B\right)\right]\\
=\frac{1}{4}\mathrm{tr}\left[\left(E_{k}B+BE_{k}\right)\left(E_{\ell}B+BE_{\ell}\right)\right]\\
=\frac{1}{4}\mathrm{tr}\left[E_{k}BE_{\ell}B+E_{k}B^{2}E_{\ell}+BE_{k}E_{\ell}B+BE_{k}BE_{\ell}\right]\\
=\frac{1}{4}\mathrm{tr}\left[\left(E_{k}BE_{\ell}\right)\left(E_{k}BE_{\ell}\right)^{*}+\left(E_{\ell}BE_{k}\right)\left(E_{\ell}BE_{k}\right)^{*}\right]\\
\geq0.
\end{multline*}

Assume that the Jordan algebra is exceptional, i.e. the Albert algebra.
There exists an $F_{4}$ automorphism of the Jordan algebra such that
$E_{k}$ is orthogonal and without loss of generality we may assume
that 
\[
E_{k}=\left(\begin{array}{ccc}
1 & 0 & 0\\
0 & 0 & 0\\
0 & 0 & 0
\end{array}\right).
\]
The idempotent $E_{\ell}$ has the form $vv*$ for some column vector
$v$ with entries in a quaternionic sub-algebra. Assume that $v=\left(\begin{array}{c}
v_{1}\\
v_{2}\\
v_{2}
\end{array}\right).$ Then
\[
E_{\ell}=\left(\begin{array}{ccc}
v_{1}\bar{v}_{1} & v_{1}\bar{v}_{2} & v_{1}\bar{v}_{3}\\
v_{2}\bar{v}_{1} & v_{2}\bar{v}_{2} & v_{2}\bar{v}_{3}\\
v_{3}\bar{v}_{1} & v_{3}\bar{v}_{2} & v_{3}\bar{v}_{3}
\end{array}\right).
\]
Now $tr\left(E_{k}\circ E_{\ell}\right)=v_{1}\bar{v}_{1}=0$ since
$E_{k}$ and $E_{\ell}$ are orthogonal. Therefore $v_{1}=0.$ Without
loss of generality we may assume that
\begin{align*}
E_{\ell} & =\left(\begin{array}{ccc}
0 & 0 & 0\\
0 & 1 & 0\\
0 & 0 & 0
\end{array}\right),\\
B & =\left(\begin{array}{ccc}
p & a & \bar{b}\\
\bar{a} & m & c\\
b & \bar{c} & n
\end{array}\right).
\end{align*}
Then
\begin{align*}
E_{k}\circ B & =\frac{1}{2}\left(\begin{array}{ccc}
2p & a & \bar{b}\\
\bar{a} & 0 & 0\\
b & 0 & 0
\end{array}\right)\\
E_{\ell}\circ B & =\frac{1}{2}\left(\begin{array}{ccc}
0 & a & 0\\
\bar{a} & 2m & c\\
0 & \bar{c} & 0
\end{array}\right)
\end{align*}
and 
\begin{multline*}
\left(E_{k}\circ B\right)\circ\left(E_{\ell}\circ B\right)=\\
\frac{1}{4}\left(\begin{array}{ccc}
2a\bar{a} & 2pa+2ma+\bar{b}\bar{c} & ac\\
2p\bar{a}+2m\bar{a}+\bar{c}\bar{b} & 2\bar{a}a & \bar{a}\bar{b}\\
\bar{c}\bar{b} & ba & 0
\end{array}\right).
\end{multline*}
Therefore 
\[
\mathrm{tr}\left[\left(E_{k}\circ B\right)\circ\left(E_{\ell}\circ B\right)\right]=\left|a\right|^{2}\geq0.
\]
\end{IEEEproof}

\subsection{Information divergence}
\begin{defn}
If the entropy is a concave function then the Bregman divergence $D_{-H}$
is called \emph{information divergence}.
\end{defn}
The information divergence is also called \emph{Kullback-Leibler divergence},
\emph{relative entropy} or \emph{quantum relative entropy}. In a Jordan
algebra we get
\begin{multline*}
D_{-H}\left(P,Q\right)=\\
-H\left(P\right)-\left(-H\left(Q\right)+\left\langle P-Q,-\nabla H\left(Q\right)\right\rangle \right)\\
=H\left(Q\right)-H\left(P\right)+\left\langle P-Q,\nabla H\left(Q\right)\right\rangle \\
=\mathrm{tr}\left[f\left(Q\right)-\mathrm{tr}\left(f\left(P\right)\right)+\mathrm{tr}\left(\left(P-Q\right)\circ f'\left(Q\right)\right)\right]\\
=\mathrm{tr}\left[f\left(Q\right)-f\left(P\right)+\left(P-Q\right)\circ f'\left(Q\right)\right]
\end{multline*}
where $f\left(x\right)=-x\ln\left(x\right).$ Now $f'\left(x\right)=-\ln\left(x\right)-1$
so that
\begin{multline*}
f\left(Q\right)-f\left(P\right)+\left(P-Q\right)f'\left(Q\right)\\
=-Q\circ\ln\left(Q\right)+P\circ\ln\left(P\right)+\left(P-Q\right)\left(-\ln\left(Q\right)-1\right)\\
=P\circ\left(\ln\left(P\right)-\ln\left(Q\right)\right)+Q-P.
\end{multline*}
Hence 
\[
D_{-H}\left(P,Q\right)=\mathrm{tr}\left[P\circ\left(\ln\left(P\right)-\ln\left(Q\right)\right)+Q-P\right]
\]
and for states $P,Q$ it reduces to 
\[
D_{-H}\left(P,Q\right)=\mathrm{tr}\left[P\circ\ln\left(P\right)-P\circ\ln\left(Q\right)\right].
\]

\begin{prop}
\label{prop:EntConv}If a state space $\mathcal{S}$ has rank 2 and
the entropy function $H$ is concave then the state space is spectral.
\end{prop}
\begin{IEEEproof}
Let $C_{-H}$ denote the capacity of the state space and. We have
\begin{align*}
C_{-H} & =\sup_{\left(q_{1},q_{2}\dots q_{n}\right)}\sum_{j}q_{i}D\left(\rho_{j}\left\Vert \sum_{i}q_{i}\rho_{i}\right.\right)\\
 & =\sup_{\left(q_{1},q_{2}\dots q_{n}\right)}H\left(\sum_{i}q_{i}\rho_{i}\right)-\sum_{j}q_{i}H\left(\rho_{j}\right)\\
 & =\sup_{\rho\in\mathcal{S}}H\left(\rho\right).
\end{align*}
If $\rho=q_{1}\rho_{1}+q_{2}\rho_{2}$ then $H\left(\rho\right)\leq-q_{1}\ln\left(q_{1}\right)-q_{2}\ln\left(q_{2}\right)\leq\ln\left(2\right).$
Therefore $\sup_{\rho\in\mathcal{S}}H\left(\rho\right)\leq\ln\left(2\right).$
Let $\bar{\sigma}$denote a capacity achieving state. Assume that
$\bar{\sigma}=p_{1}\sigma_{1}+p_{2}\sigma_{2}$ where $\sigma_{1}$
and $\sigma_{2}$ are pure orthogonal states. Then 
\begin{align*}
C_{-H} & =\sup_{\rho\in\mathcal{S}}D\left(\rho\left\Vert \bar{\sigma}\right.\right)\\
 & \geq\max_{i=1,2}\left\{ D\left(\sigma_{1}\left\Vert \bar{\sigma}\right.\right),D\left(\sigma_{2}\left\Vert \bar{\sigma}\right.\right)\right\} \\
 & =\max_{i=1,2}\left\{ \ln\left(\frac{1}{p_{1}}\right),\ln\left(\frac{1}{p_{2}}\right)\right\} \\
 & \geq\ln\left(2\right).
\end{align*}
The minimax-result them implies that $C_{-H}=\ln\left(2\right).$
Therefore $\bar{\sigma}=\frac{1}{2}\sigma_{1}+\frac{1}{2}\sigma_{2}.$
If $\rho_{1}$and $\rho_{2}$ are orthogonal and $\bar{\rho}=\frac{1}{2}\rho_{1}+\frac{1}{2}\rho_{2}.$
Then 
\begin{align*}
\ln\left(2\right) & \geq\frac{1}{2}D\left(\left.\rho_{1}\right\Vert \rho\right)+\frac{1}{2}D\left(\left.\rho_{2}\right\Vert \bar{\rho}\right)+D\left(\bar{\rho}\Vert\bar{\sigma}\right)\\
 & =\frac{1}{2}\ln\left(2\right)+\frac{1}{2}\ln\left(2\right)+D\left(\bar{\rho}\Vert\bar{\sigma}\right)\\
 & =\ln\left(2\right)+D\left(\bar{\rho}\Vert\bar{\sigma}\right).
\end{align*}
Therefore $D\left(\bar{\rho}\Vert\bar{\sigma}\right)=0$ so that $\bar{\rho}=\bar{\sigma}.$
Therefore the state space is balanced and thereby it is spectral.
\end{IEEEproof}

\subsection{Strong monotonicity}
\begin{defn}
A regret function $D_{F}$ is said to satisfy \emph{strong monotonicity}
if for any transformation $\Phi$ of the state space into itself the
equation 
\[
D_{F}\left(\Phi\left(\rho\right),\Phi\left(\sigma\right)\right)=D_{F}\left(\rho,\sigma\right)
\]
implies that there exists a recovery map $\Psi$, i.e. a map of the
state space into itself such that $\Psi\left(\Phi\left(\rho\right)\right)=\rho$
and $\Psi\left(\Phi\left(\sigma\right)\right)=\sigma$.
\end{defn}
In statistics where the state space is a simplex strong monotonicity
is well established. 
\begin{example}
Squared Euclidean distance on a spin factor is a Bregman divergence
that satisfies strong monotonicity. To see this we note an transformation
$\Phi$ can be decomposed into a translation and a linear map. Since
the transformation maps the unit ball into itself the maximal eigenvalue
of the linear map must be 1. Therefore 
\[
D_{F}\left(\Phi\left(\rho\right),\Phi\left(\sigma\right)\right)=D_{F}\left(\rho,\sigma\right)
\]
implies that $\rho$ and $\sigma$ belong to a subspace that has eigenvalue
1. The intersection of the subspace spanned by $\rho$ and $\sigma$
and the state space is a disc of radius 1 and this disc must be mapped
into another disc of radius 1 in the state space. Since any disc of
radius 1 can be mapped into any other disc of radius one there exists
a recovery map.
\end{example}
\begin{prop}
If the regret function $D_{F}$ is stronly monotone then $F$ is strictly
convex.
\end{prop}
\begin{IEEEproof}
Assume that $F$ is not strictly convex. Then there exists states
$\rho$ and $\sigma$ such that $F\left(\frac{\rho+\sigma}{2}\right)=\frac{F\left(\rho\right)+F\left(\sigma\right)}{2}.$
Then $D_{F}\left(\rho,\sigma\right)=0.$ Let $\Phi$denote a contraction
around $\frac{\rho+\sigma}{2}.$ Then $D_{F}\left(\Phi\left(\rho\right),\Phi\left(\sigma\right)\right)=0$
but a contraction cannot have a recovery map on a compact set.
\end{IEEEproof}
For density matrices over the complex numbers strong monotonicity
has been proved for completely positive maps in \cite{Jencova2006}.
Some new results on this topic can be found in \cite{Jencova2006}.

\subsection{Feasible transformations and monotonicity }

We consider a set $\mathcal{T}$ of \emph{feasible transformations}
of the state space. By a feasible transformation we mean a transformation
that we are able to perform on the state space before we choose a
feasible action. Let $\Phi:\mathcal{S}\curvearrowright\mathcal{S}$
denote a feasible transformation and let $a$ denote a feasible action.
Then $a\circ\Phi$ is the action $\sigma\to a\left(\Phi\left(\sigma\right)\right).$
Thus the set of feasible transformations acts on the set of actions.
If $\Psi$ and $\Phi$ are feasible transformations then we will assume
that $\Psi\circ\Phi$ is also feasible. Further we will assume that
the identity is feasible. Led $\mathcal{F}$ denote the monoid of
feasible transformations. Finally we will assume that $\left(1-s\right)\cdot\Psi+s\cdot\Phi$
is feasible for $s\in\left[0,1\right]$ so that $\mathcal{F}$ becomes
a convex monoid. 
\begin{prop}[The principle of lost opportunities]
If $\Phi$ is a feasible transformation then 
\begin{equation}
F\left(\Phi\left(\sigma\right)\right)\leq F\left(\rho\right).\label{eq:aftagende}
\end{equation}
 
\end{prop}
\begin{IEEEproof}
See \cite[Prop. 4]{Harremoes2017}.
\end{IEEEproof}
Since the feasible transformations increase the value of $F$ the
set of states with minimal value of $F$ is invariant under feasible
transformations. 
\begin{prop}
Let $S$ be a state space. Then the set $\mathcal{T}_{F}$ of transformations
$\Phi:\mathcal{S}\curvearrowright\mathcal{S}$ such that $F\left(\Phi\left(\sigma\right)\right)\leq F\left(\sigma\right)$
for all $\sigma\in\mathcal{S}$ is a convex monoid and for any action
$a\in\mathcal{A}_{F}$ we have that $a\circ\Phi\in\mathcal{A}_{F}.$
\end{prop}
\begin{IEEEproof}
See \cite[Prop. 5]{Harremoes2017}.
\end{IEEEproof}
\begin{cor}[Semi-monotonicity]
Let $\Phi$ denote a feasible transformation and let $\sigma$ denote
a state that minimizes the function $F$. If $D_{F}$ is a Bregman
divergence then 
\begin{equation}
D_{F}\left(\Phi\left(\rho\right),\Phi\left(\sigma\right)\right)\leq D_{F}\left(\rho,\sigma\right).\label{eq:aftagende-1}
\end{equation}
 
\end{cor}
\begin{IEEEproof}
See \cite[Cor. 1]{Harremoes2017}.
\end{IEEEproof}
Next we introduce the stronger notion of monotonicity. 
\begin{defn}
Let $D_{F}$ denote a regret function on the convex set $C.$ Then
$D_{F}$ is said to be \emph{monotone} if 
\[
D_{F}\left(\Phi\left(\rho\right),\Phi\left(\sigma\right)\right)\leq D_{F}\left(\rho,\sigma\right)
\]
for any affine transformation $\Phi:C\to C.$ 
\end{defn}
In information theory an inequality of this type is often called a
\emph{data processing inequality}. In general a regret function need
not be monotone \cite[Ex. 5]{Harremoes2017}. Recently it has been
proved that information divergence on a complex Hilbert space is decreasing
under positive trace preserving maps \cite{Mueller-Hermes2016,Christandl2016}.
Previously this was only known to hold if some extra condition like
complete positivity was assumed.
\begin{thm}
Information divergence is monotone under any positive trace preserving
map on a special Jordan algebra.
\end{thm}
\begin{IEEEproof}
The proof is a step by step repetition of the proof by M{\"u}ller-Hermes
and Reep \cite{Mueller-Hermes2016}, where they proved the theorem
for density matrices over the complex numbers. See also \cite{Christandl2016}
where the same proof technique is used. In their proof they use the
\emph{sandwiched R{\'e}nyi divergence} defined by $D_{\alpha}\left(\rho\Vert\sigma\right)=\frac{1}{\alpha-1}\ln\left(\mathrm{tr}\left[\left|\sigma^{\frac{1-\alpha}{2\alpha}}\rho\sigma^{\frac{1-\alpha}{2\alpha}}\right|^{\alpha}\right]\right),$
and we note that this quantity can be defined and manipulated as in
the proof of Reep and M{\"u}ller-Hermes as long as the algebra is associative.
\end{IEEEproof}
\begin{prop}
If a regret function is strongly monotone, then the regret function
is monotone. 
\end{prop}
\begin{IEEEproof}
Assume that the regret function $D_{F}$ is strongly monotone. Then
there exists states $\rho$ and $\sigma$ and a transformation $\Phi$
such that $D_{F}\left(\Phi\left(\rho\right)\Phi\left(\sigma\right)\right)\geq D_{F}\left(\rho,\sigma\right).$
Let $\Theta_{\epsilon}$denote a contraction around $\Phi\left(\rho\right)$
with factor $\epsilon\in\left[0,1\right].$ Then there exists a value
of $\epsilon$ such that $D_{F}\left(\Theta_{\epsilon}\left(\Phi\left(\rho\right)\right),\Theta_{\epsilon}\left(\Phi\left(\sigma\right)\right)\right)=D_{F}\left(\rho,\sigma\right).$
Therefore $\Theta_{\epsilon}\circ\Phi$ has a reverse $\Psi$ such
that 
\begin{align*}
\Psi\left(\left(\Theta_{\epsilon}\circ\Phi\right)\left(\rho\right)\right) & =\rho\\
\Psi\left(\left(\Theta_{\epsilon}\circ\Phi\right)\left(\sigma\right)\right) & =\sigma
\end{align*}
Therefore 
\begin{align*}
\left(\Phi\circ\Psi\right)\left(\Theta_{\epsilon}\left(\Phi\left(\rho\right)\right)\right) & =\Phi\left(\rho\right)\\
\left(\Phi\circ\Psi\right)\left(\Theta_{\epsilon}\left(\Phi\left(\sigma\right)\right)\right) & =\Phi\left(\sigma\right)
\end{align*}
so that $\Phi\circ\Psi$ is a recovery map of the contraction $\Theta_{\epsilon}.$
This is only possible if $\epsilon=1$ implying that $D_{F}\left(\Phi\left(\rho\right)\Phi\left(\sigma\right)\right)=D_{F}\left(\rho,\sigma\right).$
\end{IEEEproof}

\subsection{Sufficiency}

The present definition of sufficiency is based on \cite{Petz1988},
but there are a number of other equivalent ways of defining this concept.
We refer to \cite{Jencova2006} where the notion of sufficiency is
discussed in great detail.
\begin{defn}
Let $\left(\sigma_{\theta}\right)_{\theta}$ denote a family of states
and let $\Phi$ denote an affine transformation $\mathcal{S}\to\mathcal{T}$
where $\mathcal{S}$ and $\mathcal{T}$ denote state spaces. Then
$\Phi$ is said to be \emph{sufficient} for $\left(\sigma_{\theta}\right)_{\theta}$
if there exists an affine transformation $\Psi:\mathcal{T}\to\mathcal{S}$
such that $\Psi\left(\Phi\left(\sigma_{\theta}\right)\right)=\sigma_{\theta}.$
We say that $\Phi$ is \emph{reversible} if $\Phi$ is feasible and
there exist a feasible $\Psi$ such that $\Psi\left(\Phi\left(\sigma_{\theta}\right)\right)=\sigma_{\theta}.$ 
\end{defn}
\begin{prop}
If $D_{F}$ is a regret function and $\Phi$ is reversible for $\rho$
and $\sigma$ then
\[
D_{F}\left(\Phi\left(\rho\right),\Phi\left(\sigma\right)\right)=D_{F}\left(\rho,\sigma\right).
\]
\end{prop}
\begin{IEEEproof}
See \cite[Prop. 6]{Harremoes2017}.
\end{IEEEproof}
The notion of sufficiency as a property of divergences was introduced
in \cite{Harremoes2007a}. The crucial idea of restricting the attention
to transformations of the state space into itself was introduced in
\cite{Jiao2014}. It was shown in \cite{Jiao2014} that a Bregman
divergence on the simplex of distributions on an alphabet that is
not binary determines the divergence except for a multiplicative factor.
Here we generalize the notion of sufficiency from Bregman divergences
to regret functions.
\begin{defn}
We say that the regret $D_{F}$ on the state space $S$ satisfies
\emph{sufficiency} if $D_{F}\left(\Phi\left(\rho\right),\Phi\left(\sigma\right)\right)=D_{F}\left(\rho,\sigma\right)$
for any affine transformation $\mathcal{S}\to\mathcal{S}$ that is
sufficient for $\left(\rho,\sigma\right).$ 
\end{defn}
\begin{prop}
\label{prop:sufficiency}A monotone regret function $D_{F}$ satisfies
sufficiency.
\end{prop}
\begin{IEEEproof}
See \cite[Prop. 7]{Harremoes2017}.
\end{IEEEproof}
Combining the previous results we get that information divergence
on a special Jordan algebra satisfies sufficiency.
\begin{lem}
If a regret function on a state space of rank 2 satisfies sufficiency,
then the state space is spectral.
\end{lem}
\begin{IEEEproof}
For $i=1,2$ let $\sigma_{i}$ and $\rho_{i}$ denote pure states
and let $\phi_{i}$ is a test such that $\phi_{i}\left(\sigma_{i}\right)=1$
and $\phi_{i}\left(\rho_{i}\right)=0$. Let $m_{i}$ denote the midpoint
$m_{i}=\frac{1}{2}\sigma_{i}+\frac{1}{2}\rho_{i}.$ Then the transformation
$\Phi$ given by
\[
\Phi\left(\pi\right)=\phi_{1}\left(\pi\right)\sigma_{2}+\left(1-\phi_{1}\left(\pi\right)\right)\rho_{2}
\]
is sufficient for the pair $\left(\sigma_{1},m_{1}\right)$ with recovery
map $\Psi$ given by
\[
\Psi\left(\pi\right)=\phi_{2}\left(\pi\right)\sigma_{1}+\left(1-\phi_{2}\left(\pi\right)\right)\rho_{1}.
\]
Using sufficiency we have $D_{F}\left(p\sigma_{1}+\left(1-p\right)\rho_{1},m_{1}\right)=D_{F}\left(p\sigma_{2}+\left(1-p\right)\rho_{2},m_{2}\right).$
Define $f\left(p\right)=D_{F}\left(p\sigma_{1}+\left(1-p\right)\rho_{1},m_{1}\right).$
Then $D_{F}\left(p\sigma_{2}+\left(1-p\right)\rho_{2},m_{2}\right)=f\left(p\right)$
for any pair of orthogonal states $\left(\sigma_{2},\rho_{2}\right)$
so this divergence is completely determined by the spectrum $\left(p,1-p\right).$
In particular any orthogonal decomposition must have the same spectrum
so that the state space is spectral.
\end{IEEEproof}
Let $\mathcal{S}$ denote a state space of rank two with regret function
that satisfies sufficiency. Then $\mathcal{S}$ is spectral and therefore
balanced. Let $c$ denote the center of $\mathcal{S}.$ Then any state
$\sigma$ has a decomposition $\sigma=t\rho+\left(1-t\right)c$ where
$\rho$ is a pure state. We have $D_{F}\left(\sigma,c\right)=f\left(\frac{t+1}{2}\right).$
Then there exists an affine function $g$ such that $F\left(\sigma\right)=f\left(\frac{t+1}{2}\right)+g\left(t\right).$
For $t=0$ we get $g\left(0\right)=F\left(c\right).$ Therefore $F\left(\sigma\right)=f\left(\frac{t+1}{2}\right)+F\left(c\right)+h\left(t\right)$
where $h$ is a linear function that that may depend on $\rho$. 
\begin{prop}
A monotone regret function is a Bregman divergence.
\end{prop}
\begin{IEEEproof}
A regret function is given by a function $F$ and we have to prove
that this function is differentiable. Since $F$ is convex it is sufficient
to prove that $F$ is differentiable in any direction. Let $\rho$and
$\sigma$ denote two states and let $\Phi$denote a contraction around
$\rho$ by a factor $t\in\left[0,1\right]$. Then $\Phi\left(\rho\right)=\rho$and
$\Phi\left(\sigma\right)=\left(1-t\right)\cdot\rho+t\cdot\sigma.$
Let $\pi$denote a state on the line between $\rho$and $\sigma.$
Monotonicity implies that
\begin{multline*}
D_{F}\left(\Phi\left(\sigma\right),\Phi\left(\pi\right)\right)\leq D_{F}\left(\sigma,\pi\right)\\
D_{F}\left(\left(1-t\right)\cdot\rho+t\cdot\sigma,\left(1-t\right)\cdot\rho+t\cdot\pi\right)\leq D_{F}\left(\sigma,\pi\right)\\
\lim_{t\to1_{-}}D_{F}\left(\left(1-t\right)\cdot\rho+t\cdot\sigma,\left(1-t\right)\cdot\rho+t\cdot\pi\right)\leq D_{F}\left(\sigma,\pi\right)\\
\lim_{t\to1_{-}}\left(\inf_{a}F\left(\left(1-t\right)\cdot\rho+t\cdot\sigma\right)-\left\langle a,\left(1-t\right)\cdot\rho+t\cdot\pi\right\rangle \right)\\
\leq\inf_{a}F\left(\sigma\right)-\left\langle a,\sigma\right\rangle \\
F\left(\sigma\right)-\lim_{t\to1_{-}}\left(\sup_{a}\left\langle a,\left(1-t\right)\cdot\rho+t\cdot\sigma\right\rangle \right)\leq F\left(\sigma\right)-\sup_{a}\left\langle a,\sigma\right\rangle 
\end{multline*}
where $a$ on the left side should be optimal for $\Phi\left(\pi\right)$and
$a$ on the right hand side should be optimal for $\pi$. Then 
\[
\lim_{t\to1_{-}}\left(\sup_{a}\left\langle a,\left(1-t\right)\cdot\rho+t\cdot\sigma\right\rangle \right)\geq\sup_{a}\left\langle a,\sigma\right\rangle 
\]
which implies that the left derivative and the right derivative of
$F$ at $\pi$ are the same so that $F$ is differentiable at $\pi.$ 

Another way to state the result is that a regret function that is
based on a function $F$ that is not differentialbe violates monotonicity.
\end{IEEEproof}
\begin{thm}
If a state space of rank 2 has a monotone regret function, then the
state space can be represented by a spin factor.
\end{thm}
\begin{IEEEproof}
First we note that a monotone regret function is a Bregman divergence.
Since a monotone Bregman divergence satisfies sufficiency the state
space must be balanced. Let $c$ denote the center of the state space.
Without loss of generality we will assume that the state space $\mathcal{S}$
is two dimensional. Let $\Phi$denote a transformation of the state
space that first rotate it by an angle of $\theta$and then dilate
it around $c$ with a factor $r<1.$The factor can be chosen in such
a way that the boundary of $\Phi\left(\mathcal{S}\right)$ touches
the boundary of $\mathcal{S}.$ Assume that the boundary of $\Phi\left(\mathcal{S}\right)$
touches the boundary of $S$ in the pure state $\sigma'$ and let
$\phi$denote a test on $\mathcal{S}$ such that $\phi\left(\sigma'\right)=1$
and $\phi\left(c\right)=\nicefrac{1}{2}.$ Let $\Psi$ denote a map
that streches $\Phi\left(\mathcal{S}\right)$ without changing $\phi$.
The map $\Psi$ should strech so much that the boundary of $\Psi\left(\Phi\left(\mathcal{S}\right)\right)$
touch the boundary of $\mathcal{S}$ in a pure state $\rho'$ different
$\rho'$ that is not co-linear with $c$ and $\sigma'.$ Let $\rho$and
$\sigma$ denote the preimages of $\rho'$ and $\sigma'.$ Let $\pi$
denote a pure of the shortest path along the boundary of $\mathcal{S}$.
Then there exist a mixture $\bar{\sigma}=\left(1-s\right)\cdot\pi+s\cdot c$
that also has a decomposition of the form $\bar{\sigma}=\left(1-t\right)\cdot\rho+t\cdot\sigma$.

Then 
\begin{align*}
\bar{\sigma}' & =\left(1-s\right)\cdot\pi'+s\cdot c=\left(1-t\right)\cdot\rho'+t\cdot\sigma'.
\end{align*}
Using monotonicity we have 
\begin{multline*}
\left(1-t\right)\cdot D_{F}\left(\rho',\bar{\sigma}'\right)+t\cdot D_{F}\left(\sigma',\bar{\sigma}'\right)\\
\leq\left(1-t\right)\cdot D_{F}\left(\rho,\bar{\sigma}\right)+t\cdot D_{F}\left(\sigma,\bar{\sigma}\right)
\end{multline*}
so that 
\begin{multline*}
\left(1-t\right)\cdot D_{F}\left(\rho',c\right)+t\cdot D_{F}\left(\sigma',c\right)-D_{F}\left(\bar{\sigma}',c\right)\\
\leq\left(1-t\right)\cdot D_{F}\left(\rho,c\right)+t\cdot D_{F}\left(\sigma,c\right)-D_{F}\left(\bar{\sigma},c\right)
\end{multline*}
and $D_{F}\left(\bar{\sigma}',c\right)\geq D_{F}\left(\bar{\sigma},c\right).$
Since $\bar{\sigma}=\left(1-s\right)\cdot\pi+s\cdot c$ and $\bar{\sigma}'=\left(1-s\right)\cdot\pi'+s\cdot c$
and $\pi$ is a pure state $\pi'$ most also be a pure state. Therefore
$\Psi\circ\Phi$ maps pure states into pure states so that $\Psi\circ\Phi$
must be a bijection. Since any map of the state space into itself
can be streched into a bijection the state space must be a disc.
\end{IEEEproof}

\subsection{Locality}

Often it is relevant to use the following weak version of the sufficiency
property.
\begin{defn}
The regret function $D_{F}$ is said to be local if 
\[
D_{F}\left(\rho,\left(1-t\right)\rho+t\sigma_{1}\right)=D_{F}\left(\rho,\left(1-t\right)\rho+t\sigma_{2}\right)
\]
 when $\rho,\sigma_{i}$ are perfectly distinguishable for $i=1,2$
and $t\in\left]0,1\right[.$
\end{defn}
\begin{example}
On a state space of rank 2 any regret function $D_{F}$ is local.
The reason is that if $\sigma_{1}$ and $\sigma_{2}$ are states that
are orthogonal to $\rho$ then $\sigma_{1}=\sigma_{2}.$
\end{example}
\begin{prop}
A regret function $D_{F}$ that satisfies sufficiency on a convex
set is local. 
\end{prop}
\begin{IEEEproof}
\cite[Prop. 8]{Harremoes2017}.
\end{IEEEproof}
\begin{prop}
Let $C$ denote a spectral convex set with a concave entropy function
$H$. Then information divergence $D_{-H}$ satisfies locality. 
\end{prop}
\begin{IEEEproof}
Assume that $\rho$ and $\sigma$ are two perfectly distinguishable
states. Then one can make decompositions 
\begin{eqnarray*}
\rho & = & \sum_{i}p_{i}\cdot\rho_{i}\\
\sigma & = & \sum_{j}q_{j}\cdot\sigma_{j}.
\end{eqnarray*}
 Then 
\begin{align*}
D_{-H}\left(\rho,\left(1-t\right)\rho+t\sigma\right) & =\sum_{i}p_{i}\cdot\ln\frac{p_{i}}{\left(1-t\right)p_{i}}\\
 & =\sum_{i}p_{i}\cdot\ln\frac{1}{1-p}\\
 & =\ln\frac{1}{1-p},
\end{align*}
 which does not depend on $\sigma$ as long as $\rho,\sigma$ are
perfectly distinguishable.
\end{IEEEproof}
\begin{prop}
Let $C$ denote a convex set of rank 2. Then any regret function is
local. 
\end{prop}
The following lemma follows from Alexandrov's theorem. See \cite[Theorem 25.5]{Rockafeller1970}
for details.
\begin{lem}
A convex function on a finite dimensional convex set is differentiable
almost everywhere with respect to the Lebesgue measure. 
\end{lem}
\begin{thm}
\label{thm:LocEnt}Let $\mathcal{S}$ be a convex set of rank $r\geq3$
and assume that $\mathcal{S}$ is weakly spectral. If a regret function
$D_{F}$ defined on $\mathcal{S}$ is local then the state space $S$
is spectral and the regret is a Bregman divergence generated by the
entropy times some constant. 
\end{thm}
\begin{IEEEproof}
In the following proof we will assume that the regret function is
based on the convex function $F:\mathcal{S}\to\mathbb{R}.$  

Let $K$ denote the convex hull of a set $\sigma_{1},\sigma_{2},\dots\sigma_{r}$
of orthogonal states. Let $f_{i}$ denote the function $f_{i}\left(x\right)=D_{F}\left(\sigma_{i},x\sigma_{i}+\left(1-x\right)\sigma_{j}\right).$
Since $D_{F}$ is local we have $f_{i}\left(x\right)=D_{F}\left(\sigma_{i},x\sigma_{i}+\left(1-x\right)\sigma_{i+1}\right)$
for any $j\neq i.$ Note that $f_{i}$ is decreasing and continuous
from the left . Let $\rho=\sum p_{i}\sigma_{i}$ and $\sigma=\sum q_{i}\sigma_{i}.$
If $F$ is differentiable in $\rho$ then locality implies that 
\begin{eqnarray*}
D_{F}\left(\rho,\sigma\right) & = & \sum p_{i}D_{F}\left(\sigma_{i},\sigma\right)-\sum p_{i}D_{F}\left(\sigma_{i},\rho\right)\\
 & = & \sum p_{i}f_{i}\left(q_{i}\right)-\sum p_{i}f_{i}\left(p_{i}\right)\\
 & = & \sum p_{i}\left(f_{i}\left(q_{i}\right)-f_{i}\left(p_{i}\right)\right).
\end{eqnarray*}
Note that $\rho\to D_{F}\left(\rho,\sigma\right)$ is a convex function
and thereby it is continuous. Assume that $\rho_{0}$ is an arbitrary
element in $K$ and let $\left(\rho_{n}\right)_{n\in\mathbb{N}}$
denote a sequence such that $\rho_{n}\to\rho_{0}$ for $n\to\infty.$
The sequence $\left(\rho_{n}\right)_{n\in\mathbb{N}}$ can be chosen
so that regret is differentiable in $\rho_{n}$ for all $n\in\mathbb{N}.$
Further the sequence $\rho_{n}$ can be chosen such that $p_{n,i}$
is increasing for all $i\neq j.$ Then 
\begin{multline*}
D_{F}\left(\rho_{0},\sigma\right)=\sum p_{i}\left(f_{i}\left(q_{i}\right)-f_{i}\left(p_{i}\right)\right)\\
+p_{0,j}f_{j}\left(p_{0,j}\right)-p_{0,j}\lim_{n\to\infty}f_{j}\left(p_{n,j}\right).
\end{multline*}
Similarly, if the sequence $\rho_{n}$ can be chosen such that $p_{n,i}$
is increasing for all $i\neq j,j+1$ then 
\begin{multline*}
D_{F}\left(\rho_{0},\sigma\right)=\\
\sum p_{i}\left(f_{i}\left(q_{i}\right)-f_{i}\left(p_{i}\right)\right)+p_{0,j}f_{j}\left(p_{0,j}\right)-p_{0,j}\lim_{n\to\infty}f_{j}\left(p_{n,j}\right)\\
+p_{0,j+1}f_{j+1}\left(p_{0,j+1}\right)-p_{0,j+1}\lim_{n\to\infty}f_{j+1}\left(p_{n,j+1}\right),
\end{multline*}
which implies that $p_{0,j+1}f_{j+1}\left(p_{0,j+1}\right)-p_{0,j+1}\lim_{n\to\infty}f_{j+1}\left(p_{n,j+1}\right)=0$
and that $\lim_{n\to\infty}f_{j+1}\left(p_{n,j+1}\right)=f_{j+1}\left(p_{0,j+1}\right)$
for all $j$ so that 
\[
D_{F}\left(\rho_{0},\sigma\right)=\sum p_{i}\left(f_{i}\left(q_{i}\right)-f_{i}\left(p_{i}\right)\right)
\]
even if the regret is not differentiable in $\rho_{0}.$

As a function of $\sigma$ the regret has minimum when $\sigma=\rho.$
We have 
\[
x\left(f_{i}\left(y\right)-f_{i}\left(x\right)\right)+z\left(f_{j}\left(w\right)-f_{j}\left(z\right)\right)\geq0
\]
where $x+z=y+w.$ We also have 
\[
x\left(f_{j}\left(y\right)-f_{j}\left(x\right)\right)+z\left(f_{i}\left(w\right)-f_{i}\left(z\right)\right)\geq0
\]
implying that 
\[
x\left(f_{ij}\left(y\right)-f_{ij}\left(x\right)\right)+z\left(f_{ij}\left(w\right)-f_{ij}\left(z\right)\right)\geq0
\]
where $f_{ij}=\frac{f_{i}+f_{j}}{2}.$ 

Assume that $x=z=\frac{y+w}{2}$. Then
\begin{eqnarray*}
x\left(f_{ij}\left(y\right)-f_{ij}\left(x\right)\right)+x\left(f_{ij}\left(w\right)-f_{ij}\left(x\right)\right) & \geq & 0\\
f_{ij}\left(y\right)-f_{ij}\left(x\right)+f_{ij}\left(w\right)-f_{ij}\left(x\right) & \geq & 0\\
\frac{f_{ij}\left(y\right)+f_{ij}\left(w\right)}{2} & \geq & f_{ij}\left(x\right)
\end{eqnarray*}
 so that $f_{ij}$ is mid-point convex, which for a measurable function
implies convexity. Therefore $f_{ij}$ is differentiable from left
and right. We have 
\begin{multline*}
\left(y+\epsilon\right)\left(f_{ij}\left(y\right)-f_{ij}\left(y+\epsilon\right)\right)+\left(y-\epsilon\right)\left(f_{ij}\left(w\right)-f_{ij}\left(y-\epsilon\right)\right)\\
\geq0
\end{multline*}
 with equality when $\epsilon=0.$ We differentiate with respect to
$\epsilon$ from right. 
\begin{multline*}
\left(f_{ij}\left(y\right)-f_{ij}\left(y+\epsilon\right)\right)+\left(y+\epsilon\right)\left(-f_{ij+}'\left(y+\epsilon\right)\right)\\
-\left(f_{ij}\left(w\right)-f_{ij}\left(y-\epsilon\right)\right)+\left(y-\epsilon\right)\left(f_{ij-}'\left(y-\epsilon\right)\right)
\end{multline*}
 which is positive for $\epsilon=0.$ 
\begin{eqnarray*}
-y\cdot f_{ij+}'\left(y\right)+y\cdot f_{ij-}'\left(y\right) & \geq & 0\\
y\cdot f_{ij-}'\left(y\right) & \geq & y\cdot f_{ij+}'\left(y\right).
\end{eqnarray*}
 Since $f_{ij}$ is convex we have $f_{ij-}'\left(y\right)\leq f_{ij+}'\left(y\right)$
which in combination with the previous inequality implies that $f_{ij-}'\left(y\right)=f_{ij+}'\left(y\right)$
so that $f_{ij}$ is differentiable. Since $f_{i}=f_{ij}+f_{ik}-f_{jk}$
the function $f_{i}$ is also differentiable. 

We have 
\[
\frac{\partial}{\partial q_{i}}D_{F}\left(\rho,\sigma\right)=p_{i}f_{i}'\left(q_{i}\right)
\]
 and 
\[
\frac{\partial}{\partial q_{i}}D_{F}\left(\rho,\sigma\right)_{\mid\sigma=\rho}=p_{i}\cdot f_{i}'\left(p_{i}\right).
\]
 We have the condition $\sum q_{i}=1$ so using Lagrange multipliers
we get that there exist a constant $c_{K}$ such that $p_{i}\cdot f_{i}'\left(p_{i}\right)=c_{K}.$
Hence $f_{i}'\left(p_{i}\right)=\frac{c_{K}}{p_{i}}$ so that $f_{i}\left(p_{i}\right)=c_{k}\cdot\ln\left(p_{i}\right)+m_{i}$
for some constant $m_{i}.$ 

Now we get 
\begin{multline*}
D_{F}\left(\rho,\sigma\right)=\sum p_{i}\left(f_{i}\left(q_{i}\right)-f_{i}\left(p_{i}\right)\right)\\
=\sum p_{i}\left(\left(c_{K}\cdot\ln\left(q_{i}\right)+m_{i}\right)-\left(c_{K}\cdot\ln\left(p_{i}\right)+m_{i}\right)\right)\\
-c_{K}\cdot\sum p_{i}\ln\frac{p_{i}}{q_{i}}\\
=-c_{K}\cdot D_{H}\left(\rho,\sigma\right).
\end{multline*}
 Therefore there exists an affine function defined on $K$ such that
$F_{\mid K}=-c_{K}\cdot H_{\mid K}+g_{K}$. If $K$ and $L$ simplices
such that $x\in K\cap L$ then  
\[
-c_{K}\cdot H_{\mid K}\left(x\right)+g_{K}\left(x\right)=-c_{L}\cdot H_{\mid L}\left(x\right)+g_{L}\left(x\right)
\]
 so that 
\[
\left(c_{L}-c_{K}\right)\cdot H_{\mid K}\left(x\right)=g_{L}\left(x\right)-g_{K}\left(x\right).
\]
If $K\cap L$ has dimension greater than zero then the right hand
side is affine so the left hand side is affine which is only possible
when $c_{K}=c_{L}.$ Therefore we also have $g_{L}\left(x\right)=g_{K}\left(x\right)$
for all $x\in K\cap L.$ Therefore the functions $g_{K}$ can be extended
to a single affine function on the whole of $\mathcal{S}.$ 

Assume that the state $\sigma$ has two orthogonal decompositions$\sigma=\sum r_{i}\rho_{i}=\sum s_{i}\sigma_{i}.$
Since the maximum entropy state $c$ equals $\sum\frac{1}{n}\rho_{i}=\sum\frac{1}{n}\sigma_{i}$
we have that 
\begin{align*}
t\cdot\sigma+\left(1-t\right)\cdot c & =t\sum r_{i}\cdot\rho_{i}+\left(1-t\right)\cdot\sum\frac{1}{n}\rho_{i}\\
 & =\sum\left(t\cdot r_{i}+\left(1-t\right)\cdot\frac{1}{n}\right)\rho_{i}\\
 & =\sum\left(t\cdot s_{i}+\left(1-t\right)\cdot\frac{1}{n}\right)s_{i}.
\end{align*}
Therefore 
\begin{multline*}
H\left(t\cdot\sigma+\left(1-t\right)\cdot c\right)\\
=-\sum_{i=1}^{n}\left(t\cdot r_{i}+\left(1-t\right)\cdot\frac{1}{n}\right)\ln\left(t\cdot r_{i}+\left(1-t\right)\cdot\frac{1}{n}\right)\\
=-\sum_{i=1}^{n}\left(t\cdot s_{i}+\left(1-t\right)\cdot\frac{1}{n}\right)\ln\left(t\cdot s_{i}+\left(1-t\right)\cdot\frac{1}{n}\right).
\end{multline*}
Assume $r_{1}\leq r_{2}\leq\dots\leq r_{n}$ and $s_{1}\leq s_{2}\leq\dots\leq s_{n}.$
The analytic continuation of $t\to\sum_{i=1}^{n}\left(t\cdot r_{i}+\left(1-t\right)\cdot\frac{1}{n}\right)\ln\left(t\cdot r_{i}+\left(1-t\right)\cdot\frac{1}{n}\right)$
is defined for values of $t$ between $\frac{1}{1-nr_{n}}$ and $\frac{1}{1-nr_{1}}$
and the analytic continuation of $t\to\sum_{i=1}^{n}\left(t\cdot s_{i}+\left(1-t\right)\cdot\frac{1}{n}\right)\ln\left(t\cdot s_{i}+\left(1-t\right)\cdot\frac{1}{n}\right)$
is defined for values of $t$ between $\frac{1}{1-ns_{n}}$ and $\frac{1}{1-ns_{1}}.$
Since the functions $t\to\sum_{i=1}^{n}\left(t\cdot r_{i}+\left(1-t\right)\cdot\frac{1}{n}\right)\ln\left(t\cdot r_{i}+\left(1-t\right)\cdot\frac{1}{n}\right)$
and $t\to\sum_{i=1}^{n}\left(t\cdot s_{i}+\left(1-t\right)\cdot\frac{1}{n}\right)\ln\left(t\cdot s_{i}+\left(1-t\right)\cdot\frac{1}{n}\right)$
are identical we most have $\frac{1}{1-nr_{n}}=\frac{1}{1-ns_{n}}$
and $s_{n}=r_{n}.$ Therefore 
\begin{multline*}
\sum_{i=1}^{n-1}\left(t\cdot r_{i}+\left(1-t\right)\cdot\frac{1}{n}\right)\ln\left(t\cdot r_{i}+\left(1-t\right)\cdot\frac{1}{n}\right)\\
=\sum_{i=1}^{n-1}\left(t\cdot s_{i}+\left(1-t\right)\cdot\frac{1}{n}\right)\ln\left(t\cdot s_{i}+\left(1-t\right)\cdot\frac{1}{n}\right)
\end{multline*}
 and this argument can be repeated to show that $r_{i}=s_{i}$ that
that the two spectra are identical.
\end{IEEEproof}
Combining Theorem \ref{thm:LocEnt} with Proposition \ref{prop:EntConv}
leads to the following result.
\begin{cor}
A weakly state space with a strichtly concave entropy function is
spectral.
\end{cor}
In \cite{Krumm2016} it was proved that if a state space is symmetric
and spectral then the entropy function is concave.

\section{Conclusion}

In this paper we have demonstrated that state spaces are spectral
sets if a well-behaved divergence function can be defined. The simple
Jordan algebras are symmetric spectral sets and we conjecture that
all symmetric spectral sets can be represented on Jordan algebras.
A complete classification of spectral sets is highly desirable but
does not exist yet.

\section*{Acknowledgment}

The author want to thank Prasad Santhanam for inviting me to the Electical
Engineering Department, University of Hawai\textquoteleft i at M\={a}noa,
where some of the ideas presented in this paper were developed. I
also want to thank Jan Naudts, Alexander M{\"u}ller-Hermes, and Chris
Perry for useful discussions.

\end{document}